\documentclass[3p,onecolumn]{elsarticle}

\usepackage{hyperref}
\usepackage{txfonts}
\usepackage{natbib}
\usepackage{graphicx}
\usepackage[utf8]{inputenc}

\journal{Icarus}

\biboptions{authoryear}

\def\orten{2007\,OR\ensuremath{_{10}}}
\def\ortenlong{(225088)~2007\,OR\ensuremath{_{10}}}

\def\ggg{G!k\'un$||$'h\`omd\'im\`a}

\def\geomalb{p$_{\mathrm{V}}$}
\def\absmag{H$_{\mathrm{V}}$}
\def\gcc{g\,cm$^{-3}$}
\newcommand{\tiunit}{$\mathrm{J\,m^{-2}\,s^{-1/2}\,K^{-1}}$}
\newcommand{\deff}{$\mathrm{D_{eff}}$}

\begin{document}

\newcommand{\arcmin}{$^{\prime}$}
\newcommand{\arcsec}{$^{\prime\prime}$}
\newcommand{\degr}{$^{\circ}$}
\newcommand{\mjysr}{MJy\,sr$^{-1}$}
\newcommand{\dr}{2012\,DR$_{30}$}
\newcommand{\angstrom}{\mbox{\normalfont\AA}}
\newcommand\sun{\hbox{$\odot$}}
\def\lesssim{\mathrel{\hbox{\rlap{\hbox{\lower3pt\hbox{$\sim$}}}\hbox{\raise2pt\hbox{$<$}}}}}
\def\degr{\hbox{$^\circ$}}
\def\arcmin{\hbox{$^\prime$}}
\def\arcsec{\hbox{$^{\prime\prime}$}}
\def\utw{\smash{\rlap{\lower5pt\hbox{$\sim$}}}}
\def\udtw{\smash{\rlap{\lower6pt\hbox{$\approx$}}}}
\def\fd{\hbox{${.}\!\!^{\rm d}$}}
\def\fh{\hbox{${.}\!\!^{\rm h}$}}
\def\fm{\hbox{${.}\!\!^{\rm m}$}}
\def\fs{\hbox{${.}\!\!^{\rm s}$}}
\def\fdg{\hbox{${.}\!\!^\circ$}}
\def\farcm{\hbox{${.}\mkern-4mu^\prime$}}
\def\farcs{\hbox{${.}\!\!^{\prime\prime}$}}
\def\fp{\hbox{${.}\!\!^{\scriptscriptstyle\rm p}$}}

\sloppy

\begin{frontmatter}
\title{The mass and density of the dwarf planet \ortenlong}


\author[csfk]{Csaba~Kiss} 
\author[csfk]{G\'abor~Marton} 
%
%
%
\author[swri]{Alex H. Parker}
%
\author[lowel]{Will Grundy}
%
\author[csfk,elte]{Anik\'o Farkas-Tak\'acs}
%
\author[stsci]{John Stansberry}
%
\author[csfk,elte]{Andr\'as P\'al}
\author[mpe]{Thomas M\"uller}
\author[goddard]{Keith S. Noll}

\author[gemini]{Megan E. Schwamb}
\author[psi]{Amy C. Barr}
\author[swri]{Leslie A. Young}
%
\author[csfk]{J\'ozsef Vink\'o}
%
\address[csfk]{Konkoly Observatory, Research Centre for Astronomy and Earth Sciences, Hungarian Academy of Sciences, Konkoly Thege 15-17, H-1121~Budapest, Hungary}
\address[swri]{Southwest Research Institute, Boulder, Colorado, USA}
\address[lowel]{Lowell Observatory, Flagstaff, Arizona, USA}
\address[elte]{E\"otv\"os Lor\'and University, P\'azm\'any P. st. 1/A, 1171 Budapest, Hungary}
\address[stsci]{Space Telescope Science Institute, 3700 San Martin Dr., Baltimore, MD 21218, USA}
\address[mpe]{Max-Planck-Institut f\"ur extraterrestrische Physik, Giesenbachstrasse, Garching, Germany}
\address[goddard]{NASA Goddard Space Flight Center, Greenbelt, MD 20771, USA}
\address[gemini]{Gemini Observatory, Northern Operations Center, 670 North A'ohoku Place, Hilo, Hawaii 96720, USA}
\address[psi]{Planetary Science Institute, 1700 E. Ft. Lowell, Suite 106, Tucson, AZ 85719, USA}




\begin{abstract}

The satellite of \ortenlong\, was discovered on archival Hubble Space Telescope images 
and along with new observations with the WFC3 camera in late 2017 we have been able to determine the orbit. The orbit's notable eccentricity, $e$\,$\approx$\,0.3, may be a consequence of an intrinsically eccentric orbit and slow tidal evolution, but may also be caused by the Kozai mechanism. Dynamical considerations also suggest that the moon is small, \deff\,$<$\,100\,km.
Based on the newly determined system mass of 1.75\,$\cdot$10$^{21}$\,kg, \orten\, is the fifth most massive dwarf planet after Eris, Pluto, Haumea and Makemake.  
The newly determined orbit has also been considered as an additional option in our radiometric analysis, provided that the moon orbits in the equatorial plane of the primary. {Assuming a spherical shape for the primary} this approach provides a size of {1230$\pm$50\,km}, with a slight dependence  on the satellite orbit orientation and primary rotation rate chosen, and a bulk density of {1.75$\pm$0.07\,g\,cm$^{-3}$ } for the primary. A previous size estimate that assumed an equator-on configuration (1535$^{+75}_{-225}$\,km) would provide a density of 0.92$_{-0.14}^{+0.46}$\,g\,cm$^{-3}$, unexpectedly low for a 1000\,km-sized dwarf planet. 
\end{abstract}

\end{frontmatter}

\section{Introduction} \label{sect:intro}

Satellites are very important in studying the formation and evolution of Kuiper belt objects \citep[see][for a summary]{Noll2008}. The orbit of a satellite allows us to obtain accurate system mass and also density when the size of the main body is known (typically from radiometry or occultation measurements). 
Densities are also indicative of the internal structure, and are important constraints for satellite formation theories. It is possible that systems with small and large moons formed by different processes. Systems with large moons may have formed in low-velocity grazing collisions, both bodies retaining their original compositions and also the primordial densities. 
Systems with small moons may have formed in collisions when low-density icy material is lost, increasing the bulk density of the primary \citep{Barr2016}. 

The satellite orbits of most large KBO binaries are nearly circular. An exception is (50000) Quaoar, where the orbit of Weywot is moderately eccentric ($\epsilon$\,=\,0.14), an orbital state that is likely not the consequence of a tidal evolution from an initially circular orbit. The long orbit evolution timescale obtained for Weywot indicates instead that it may have formed with a non-negligible eccentricity \citep{Fraser2013}. 

The satellite of \ortenlong\, (hereafter shortened to \orten) was discovered on archival images obtained with the WFC3 camera of the Hubble Space Telescope \citep{Kiss2017}. 
This discovery completes the list of outer solar system dwarf planets with known satellites: now all bodies larger than $\sim$1000\,km in diameter are known to harbor moons (Pluto-Charon, Eris, Haumea, Makemake, Quaoar, Orcus). The existence of a satellite was originally suspected from the long rotation period ($\sim$44.8\,h) derived from a Kepler-K2 multi-day light curve \citep{Pal2016}. The initial discovery was based on observations at two epochs only, therefore the orbit of the satellite could not be derived unambiguously from these data alone.   

Here we report on successful recovery observations of the satellite of \orten, taken with the WFC3 camera of the Hubble Space Telescope (HST) in 2017. The observations allow us to determine the orbit sufficiently well to obtain system mass and estimate the density of the primary. We also give a short assessment of possible orbital evolution and the consequences for both the primary and the satellite. 

\begin{figure}[ht!]
\begin{center}
\includegraphics[width=0.47\textwidth]{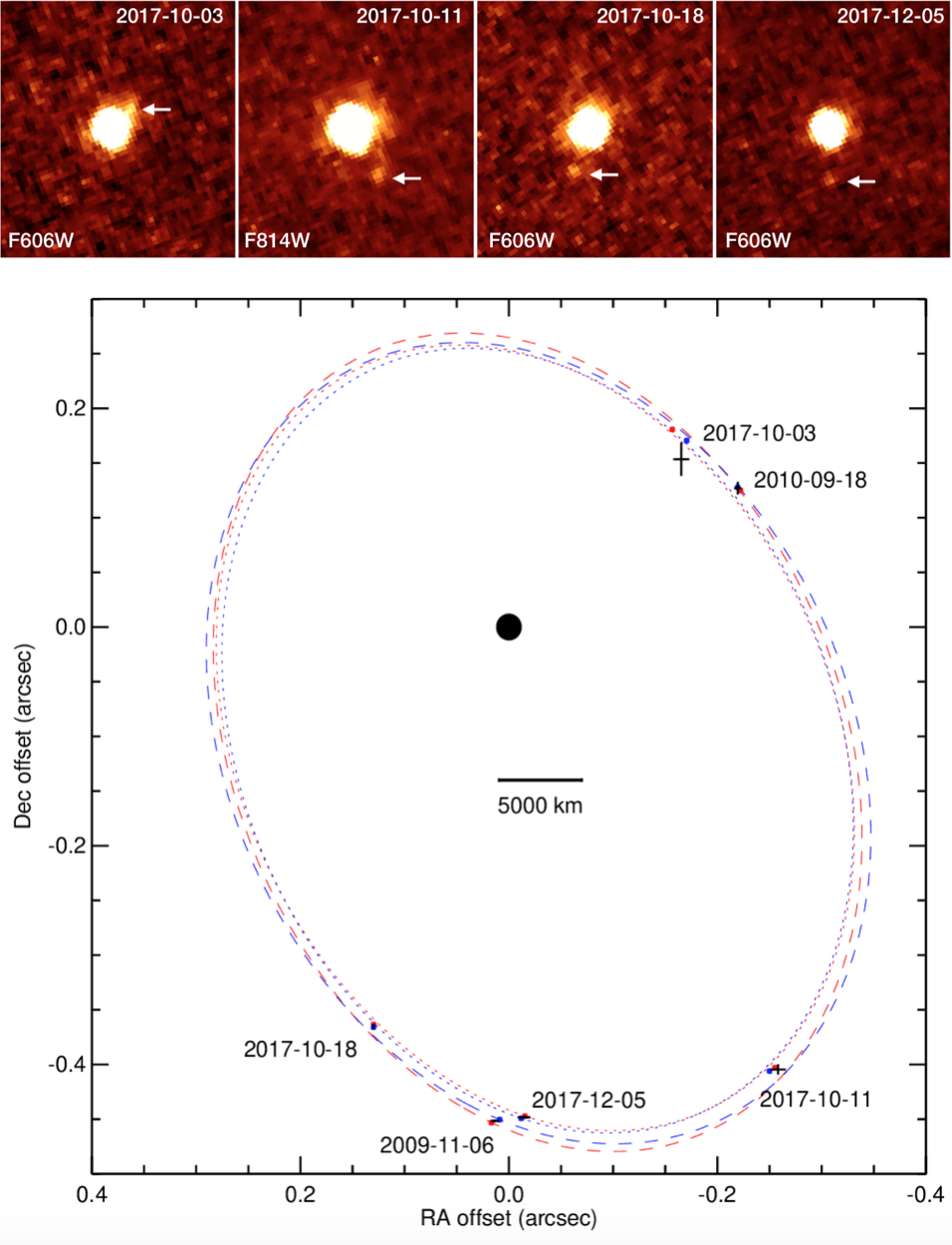} 
\end{center}
\caption{\textit{Upper panel:} Hubble Space Telescope WFC3/UVIS images of \orten, 
obtained in October-December 2017 recovery observations (see Table~\ref{table:astrometry} for details). 
\textit{Lower panel:} Sky-projected orbit of the satellite around \orten. Dashed lines correspond to the orbit at the time of the 2009 observation and the dotted ones are at the time of the recovery observations in 2017.  The blue ellipse corresponds to the prograde, red one to the retrograde solution.  
The points with error bars mark the observed positions of the satellite (see Table~\ref{table:astrometry}) while the small blue and red points mark the expected relative positions at the time of the observations, derived from the orbital solutions. The points marking the satellite are sized to 100\,km radius, while that of \orten{} (in the center) corresponds to a diameter of 1535\,km.
\label{fig:images}}
\end{figure}

\section{Observations and data analysis} \label{sect:obsanddata}


New observations of \orten{} were obtained with HST in the framework of the proposal "The Moons of Kuiper Belt Dwarf Planets Makemake and \orten" (proposal ID: 15207, PI: A.H.~Parker) at four epochs in October and December, 2017 (see Table~\ref{table:astrometry}). The WFC3/UVIS camera system with the UVIS2-C512C-SUB aperture was used to take multiple exposures, alternating between the F350LP and either the F606W or the F814W filters. 
We created co-added images in the co-moving frame of \orten{} using images obtained with the same filters. The satellite was clearly visible and well-separated from \orten{} on the images taken on October 10, 18 and December 5, but was quite close to the bright primary on October 3. We used point-spread function (PSF) subtracted images to perform astrometry and photometry of the satellite, using the same DAOPHOT-based routines as in \citet{Kiss2017}. The model PSFs used for subtraction were created using the TinyTim \citep{Krist2010} software, using specific setups (date, camera system, target’s pixel position, focal length). 
 The results are summarized in Table~\ref{table:astrometry} and Fig.~\ref{fig:images}. 
 { Due to the proximity of the satellite and the primary, the astrometry of the 2017 October 3 measurement has a notably higher uncertainty than the other measurements. While in the other cases the images with and without PSF-subtraction provided nearly identical astrometry ($<$1\,mas), these differences are an order of magnitude larger at the October 3 epoch ($>$10\,mas) that is also reflected in the quoted astrometric uncertainties. 
   }


\section{Orbit fitting}

After the October 2017 observations, we generated a collection of thousands of orbits consistent with the ensemble of astrometric data, using Monte Carlo procedures \citep{Grundy2008}.  This cloud of orbits provided a representation of the probability distribution in orbital element space. It contained a number of dense clumps corresponding to distinct orbit solutions differing in their orbital periods, eccentricities, semi-major axes, etc.  
Each clump was used to provide initial parameters for a least-squares fit, using the Ameoba downhill simplex algorithm \citep[][]{Nelder,Press1992} to adjust the orbital elements to minimize the residuals between observations and predicted positions.
We chose December 2017 as the optimal time for the last observation because the cloud of possible orbits was well dispersed, but not homogeneous, when projected on the sky plane at that epoch.
After completion of the final observation, the Monte Carlo orbit fits were repeated, and all but one of the clumps of potential orbits were rejected, leaving only the solution in Table~\ref{table:orbitsolutions} and also illustrated in Fig.~\ref{fig:images}. 


This pole solution has two counterparts, mirrors of one another through the sky plane at the time of the 2017 observations.  To distinguish which of the prograde and retrograde solutions is the correct one will require waiting for \ortenlong{} to move further along its heliocentric orbit, enabling Earth-based observers to view the system from a different direction.  But already, the period and semi-major axis are reasonably well determined, enabling us to derive the system mass.  Additionally, the eccentricity is significantly non-zero  ($e$\,=\,0.29 and 0.28 in the prograde and retrograde cases, respectively), \rm a result that we explore in more detail below.  The prograde and retrograde solutions provide a mean mass estimate of 1.75$\pm$0.07$\cdot$10$^{21}$\,kg, which is the fourth largest known mass among dwarf planets after Eris, Pluto and Haumea \citep[][respectively]{Stern2015,Brown+Schaller,Ragozzine2009}. This mass is very similar to that of Charon \citep[1.586$\cdot$10$^{21}$\,kg][]{Nimmo2017}.

The small residuals of the individual astrometry points (Fig.~\ref{fig:images}) and the observed change between the apparent orbits of the satellite in the first (2009/10) and second (2017) observing seasons agrees well with the assumption that the binary orbit is stable, i.e. the orbit pole did not change between the two observational seasons, and the change of the apparent orbit of the satellite can be explained by the aspect angle change due to the displacement of \orten\, on its heliocentric orbit. The largest, 1.6\,$\sigma$ residual is between the model and observed positions of the October 3, 2017 measurement; in the other cases it is $\lesssim$1\,$\sigma$.  

\rm


\section{Photometry results and colors}

Based on the differential photometry of \orten\, and the satellite (see Table~\ref{table:astrometry}) we obtained average brightness differences of $\Delta(F606W)$\,=\,4\fm68$\pm$0\fm11 (observations on October 3, 18, and December 5) and $\Delta(F814W)$\,=\,5\fm01$\pm$0\fm30 (October 11). We use a system-integrated absolute brightness of H$_V$\,=\,2.34$\pm$0.01 and the color V--I\,=\,1.65\,$\pm$0.03 \citep{B14} to obtain absolute brightness values for the satellite from the relative photometry. When transforming the HST/WFC3 photometry to the Johnson-Cousins system (F606W to V and F814W to I) we applied a V band correction of 0\fm10 due to the color difference of the satellite and the primary; in the case of the I-band brightness values the correction was much smaller ($<$0\fm003) 
\citep[see][for the transformations between the HST/WFC3 and the Johnson-Cousins photometric systems]{Sahu}. For the absolute brightness and color of the satellite we obtained H$_V^{s}$\,=\,6.93$\pm$0.15 and (V--I)$_s$\,=\,1.22$\pm$0.17, i.e. it is somewhat less red than the primary. From this color a spectral slope of S$^\prime_s$\,=\,19$\pm$7\%/(1000\,\AA)\,can be derived, while the spectral slope of the notably redder primary is S$^\prime_p$\,=\,42$\pm$2\%/(1000\,\AA). 
 
Mid-sized trans-Neptunian binaries typically have nearly equal colors \citep{Benecchi2009}, indicating that in most cases the satellite co-formed in a locally homogeneous, but globally heterogeneous protoplanetary disk. 
For larger bodies, however, color differences of 0\fm2--0\fm3 are common, as it is the case for Pluto-Charon \citep{Grundy2016}, Eris-Dysnomia \citep{Brown+Schaller} or Orcus-Vanth \citep{Brown2010}. In the latter two cases the primaries have nearly solar colors and the satellites are redder and  darker, however, in these cases the colors of the primaries may not be original. The \orten{} system seems to have the largest color difference among trans-Neptunian binaries, with $\Delta$(V-I)\,=\,0.43$\pm$0.17. 


\begin{table*}
\caption{Relative astrometry (J2000) and photometry of the satellite with respect to \orten.  The first two lines correspond to the discovery epochs \citep[see][]{Kiss2017}, the next four lines represent the results of the HST recovery observations in 2017.  $\delta$r$_p$ and $\delta$r$_r$ are the astrometry residuals (R.A. and DEC combined) from the best fit prograde and retrograde model as presented in Table~\ref{table:orbitsolutions} and Fig~\ref{fig:images}. Note that the HST/WFC3 pixel scale is $\sim$40\,mas. }


\begin{center}
\begin{tabular}{cccccc|cc}
\hline
MJD start   &  MJD end    & $\Delta\alpha$ & $\Delta\delta$ & filter & $\Delta$m & $\delta$r$_p$ & $\delta$r$_r$\\
            &             &  (mas)         &    (mas)       &        & (mag)     & \multicolumn{2}{c}{(mas)}    \\
\hline
55141.71282 & 55141.72189 &  +13$\pm$4 & -452$\pm$2 & F606W/F814W & 4.42$\pm$0.21 / 4.35$\pm$0.25 &  5  &  4 \\
55457.66116 & 55457.66987 & -219$\pm$3 & +127$\pm$6 & F606W/F775W & 4.15$\pm$0.13 / 4.43$\pm$0.30 &  1  &  4 \\
\hline
58029.66532 & 58029.68957 & -165$\pm$8 &  +153$\pm$15 & F606W & 4.93$\pm$0.30 & 18 & 28 \\
58037.34117 & 58037.36635 & -258$\pm$7 &  -405$\pm$5  & F814W & 5.01$\pm$0.15 &  8 & 3 \\
58044.22969 & 58044.25393 & +130$\pm$2 &  -365$\pm$3  & F606W & 4.65$\pm$0.15 &  1 & 2 \\
58092.23763 & 58092.26689 &  -15$\pm$6 &  -448$\pm$2  & F606W & 4.64$\pm$0.17 &  3 & 2 \\
\hline
\end{tabular}
\end{center}
\label{table:astrometry}
\end{table*}

\begin{table}[ht!]
\caption{Orbital solutions and derived parameters from the HST observations. The orbital elements correspond to the epoch of 2457000.0 (JD). }
\begin{center}
\begin{tabular}{ccc}
\hline
            &    prograde    &    retrograde   \\
\hline                
P (day)    &       25.22073$\pm$0.000357   &    25.22385$\pm$0.000362 \\
a (km)     &       24021$\pm$202           &    24274$\pm$193         \\
e          &      0.2908$\pm$0.0070        &    0.2828$\pm$0.0063     \\
i (deg)    &       83.08$\pm$0.86          &    119.14$\pm$0.89       \\
$\epsilon$ (deg) &  205.57$\pm$0.95        &    294.47$\pm$1.38       \\
$\Omega$   (deg) &   31.99$\pm$1.07        &    104.09$\pm$0.82       \\
$\varpi$   (deg) &  109.05$\pm$1.88        &    199.15$\pm$1.67       \\
\hline
 M$_{sys}$ (kg)        &      (1.726$\pm$0.043)$\cdot$10$^{21}$  & (1.781$\pm$0.043)$\cdot$10$^{21}$   \\
$\alpha_{pole}$ (deg)  &    301.990$\pm$1.021   &     14.096$\pm$0.679 \\
$\delta_{pole}$ (deg)  &      6.914$\pm$0.451   &    -29.143$\pm$0.408 \\
$\lambda_{pole}$ (deg) &  305.972$\pm$1.160 &  0.098$\pm$0.723 \\
$\beta_{pole}$ (deg)   &  26.447$\pm$0.550  & -32.101$\pm$0.516 \\
i$_{helio}$ (deg)      &  51.828$\pm$0.829  &   129.050$\pm$0.703 \\
%
\hline
\end{tabular}
\end{center}
\label{table:orbitsolutions}
\end{table}
\section{Radiometric size estimates \label{Sect:Radiometry}}

The thermal emission of \orten{} was observed with the PACS camera of the Herschel Space Observatory, and these data were 
analysed in detail in \citet{Pal2016}. Both the Near-Earth Thermal Asteroid Model and the thermophysical model (TPM) pointed to a same best-fit size of 1535$_{-225}^{+75}$\,km. In that paper two TPM configurations were tested: a pole-on and an equator-on, and the latter one gave the best fit to the observed flux densities. 
Although the recent HST observations do not constrain the rotation axis orientation directly, one may assume that the orbit of the satellite is in the equatorial plane of \orten and use our two pole orientations for the spin axis of the primary. 
Overall, we considered { four} possible pole orientations, presented in Table~\ref{table:tpms}. 
We allowed  thermal inertias in the range of $\Gamma$\,=\,0.1--50\,\tiunit, and a constant emissivity of $\epsilon$\,=\,0.9 in the TPM models. As was recently demonstrated \citep{Fornasier2013,Lellouch2017} far-infrared and submillimetre flux densities of outer solar system objects may be affected by lower-than-unity relative emissivities with respect to those in the mid-infrared regime. While this is most expressed in the submillimetre, a slight deviation in relative emissivity ($\epsilon_{rel}$\,$\approx$\,0.9) was also observed at 160\,$\mu$m for some objects. If the emissivity of \orten\, were depresssed at 160\,$\mu$m that would affect our derived diameter; however, the data show no indication of such an effect for \orten.

{ Due to its large size it is not expected that the shape of \orten\, would deviate significantly from a sphere. For a rotating body with relatively low angular velocity the expected shape is a Maclaurin spheroid with semi-major axes $a$\,=\,$b$\,$>$\,$c$, and rotation around the $c$ axis \citep{Plummer1919}. The flattening ($\epsilon$\,=\,$1-c/a$) can be calculated for a specific normalized angular velocity of $\omega^2/\pi G \rho$, and for $\rho$\,=\,1\,\gcc\, we obtained $\epsilon$\,=\,0.03 and $\epsilon$\,=\,0.007 for P\,=\,22.4\,h and 44.8\,h, respectively. This is very far from the fast rotator cases when the equlibrium configuration is a Jacobi ellipsoid. As $a$\,=\,$b$ for a Maclaurin ellipsoid, this shape results in a flat light curve. 

Charon has a mass very similar to that of \orten, m$_{Ch}$\,=\,1.586$\cdot$10$^{21}$\,kg, and its shape is very close to a sphere, with $a$\,=\,606$\pm$1\,km and flattening $<$0.5\% \citep{Nimmo2017}, despite the presence of Pluto. This suggests that the observed light curve is likely caused by surface features (albedo variegations) rather than by a distorted shape in the case of \orten, too. Therefore we consider a sphere for \orten\, in the thermal modelling as the main shape option. 

However, one may eventually assume that we see a distorted body with $a$\,$>$\,$b$\,$>$\,$c$ that leads to the observed light curve. A tidally distorted body would have $(a-b)$\,=\,$4(b-c)$; for the equator-on and prograde/retrograde equatorial satellite cases the observed light curve amplitude of $\Delta m$\,=\,0\fm09 \citep{Pal2016} requires $b$\,=\,0.92 and $b$\,=\,0.74 (Cases 4 and 5a...d). For the thermal emission here we assume that we observed \orten\, at a 'mean' rotational phase. In these cases the estimated effective diameters are different from those in the corresponding spherical cases due to the different projected area, leading to different effective diameters and densities as well.  

} 
\begin{table*}
\caption{Thermophysical models setups with different rotational axis orientations, represented by the ecliptic coordinates of the rotational pole ($\lambda_p,\beta_p$), and by the subsolar latitude $\beta_{ss}$. We also show the corresponding best fit TPM solution of effective diameter \deff, geometric albedo \geomalb, thermal inertia $\Gamma$ and the density of the primary derived from these values. Comments: eon -- equator-on; pon -- pole-on; pg/rg -- prograde/retrograde satellite in the equatorial plane of the primary;  s -- spherical; e -- tidally distorted ellipsoid}
\begin{tabular}{cccccc|cccc}
\hline
Case &$\lambda_p$ & $\beta_p$ & $\beta_{ss}$ & P$_\mathrm{rot}$ & comment & \deff & \geomalb & $\Gamma$ & $\rho$ \\ 
    &  (deg)      &  (deg)    &   (deg)       & (h) &         &  (km) &          & (\tiunit)& (g\,cm$^{-3}$)\\
\hline         
1 & 331.9 & 86.7 & 0 & 44.8 & eon/s &1531&0.09& 3 & $0.92^{+0.46}_{-0.14}$\\
2 & 331.9 & -3.3 & 90 & -- & pon/s  & $1158\pm 32$& 0.16$\pm$0.01 & unconstrained&$2.15\pm 0.17$ \\
\hline
3a & 306.0 &  26.4 & 51 & 44.8 & pg/s & 1224$\pm$55 & 0.14$\pm$0.01 & 1--5 & 1.80$\pm$0.16 \\
3b & 306.0 &  26.4 & 51 & 22.4 & pg/s & 1238$\pm$50 & 0.14$\pm$0.01 & 1--5 & 1.74$\pm$0.16 \\
3c & 0.1   & -32.1 & 51 & 44.8 & rg/s & 1227$\pm$56 & 0.14$\pm$0.01 & 1--5 & 1.79$\pm$0.16 \\
3d & 0.1   & -32.1 & 51 & 22.4 & rg/s & 1241$\pm$50 & 0.14$\pm$0.01 & 1--5 & 1.73$\pm$0.16 \\
\hline
4  & 331.9 & 86.7 & 0 & 44.8 & eon/e & 1549 & 0.09 & 3 & $0.89^{+0.44}_{-0.14}$\\
\hline
5a & 306.0 &  26.4 & 51 & 44.8 & pg/e & 1155$\pm$52 & 0.16$\pm$0.01 & 1--5 & 2.13$\pm$0.17 \\
5b & 306.0 &  26.4 & 51 & 22.4 & pg/e & 1169$\pm$47 & 0.16$\pm$0.01 & 1--5 & 2.07$\pm$0.17 \\
5c & 0.1   & -32.1 & 51 & 44.8 & rg/e & 1158$\pm$53 & 0.16$\pm$0.01 & 1--5 & 2.13$\pm$0.17 \\
5d & 0.1   & -32.1 & 51 & 22.4 & rg/e & 1172$\pm$47 & 0.15$\pm$0.01 & 1--5 & 2.05$\pm$0.17 \\
\hline
\end{tabular}
\label{table:tpms}
\end{table*}
{ As input for the thermopysical model calculations we used the 70, 100 and 160\,$\mu$m flux densities presented in \citet{Pal2016}, a rotation period of P$_{\mathrm rot}$\,=\,44.81\,h or the half period, P\,=\,22.4\,h \citep{Pal2016}, a low to intermediate surface roughness (0.1--0.3 r.m.s. of surface slopes), and an absolute magnitude of H$_V$\,=\,2\fm34$\pm$0\fm05 \citep{B14}.}

As was shown in \citet{Kiss2017} the satellite can noticeably contribute to the thermal emission only if its surface is very dark (\geomalb\,$<$\,4\%). As we argue later in the this paper, dynamical considerations strongly favour a small satellite with \geomalb\,$>$\,20\%\, therefore the satellite's contribution is negligible in the thermal emission models.  

The best fit to the data is given by the Case-1 (equator-on, subsolar latitude of $\beta_{ss}$\,$\approx$\,0\degr) configuration (reduced $\chi^2$\,$\approx$\,0.1), resulting in $\Gamma$\,=\,2--6\,\tiunit, with an optimum solution of $\Gamma$\,=\,3\,\tiunit, \deff\,=\,1531\,km, and p$_V$\,=\,0.09. 

\citet{Lellouch2013} obtained $\Gamma$\,=\,2.5$\pm$0.5\,\tiunit\, for typically 100\,km-sized objects observed at heliocentric distances of r$_h$\,=\,20--50\,AU. At the distance of \orten{} the thermal inertia of a similar surface would be lower due to the lower surface temperatures: assuming that the T$^3$ term dominates in the thermal conductivity, thermal inertia scales 
as $\propto$\,r$_h^{-3/4}$ \citep{Delbo2015}, i.e. a $\Gamma$ of a factor of $\sim$2 lower is expected at the distance of \orten. 

The thermal inertia of larger bodies, however, may be notably larger. 
\citet{Lellouch2011,Lellouch2016} obtained $\Gamma_{Pl}$ = 16-26\,\tiunit and $\Gamma_{Ch}$\,=\,9--14\,\tiunit, for Pluto and Charon, respectively. 
These high $\Gamma$ values are thought to be caused by the slow rotation, and the $\propto$\,P$^{1/2}$ dependence of the diurnal skin depth on the rotation period. \orten\, rotates faster than Pluto and Charon (6.38\,d orbital/rotation period of Pluto-Charon versus 44.\,8h), but still much slower than a typical trans-Neptunian object (P\,=\,$\sim$6--12\,h). 

For \orten\, this suggests a factor of $\sim$2 reduction of $\Gamma$ compared with Pluto or Charon, altogether a factor of $\sim$4 smaller values, considering the r$_h$ dependence as well. This gives $\Gamma$\,=\,4--6\,\tiunit\, for 'Pluto-like', and $\Gamma$\,=\,2.2--3.5\,\tiunit\, for 'Charon-like' surfaces, in a very good agreement with that obtained from the thermophysical model analysis assuming Case-1 (equator-on). The analysis of the thermal emission of a Haumea also indicates a similarly high thermal inertia \citep{Muller2018}.

The pole-on configuration (Case~2, $\beta_{ss}$\,$\approx$\,90\degr) provides a size of \deff\,=\,1158$\pm$32\,km, and this value is independent of the thermal inertia chosen. 
\citet{Pal2016} found a peak-to-peak light curve amplitude of $\Delta$m\,=\,0\fm09, which in the case of a near-to-pole-on configuration would be a significantly supressed fraction of a much larger intrinsic light curve amplitude that would be seen at low obliquity. $\Delta$m\,=\,0\fm09 associated with $\beta_{ss}$\,$>$\,80\degr\, would require an extremely elongated body (not expected due to the slow rotation) or a very high (a factor of $\sim$2 or larger) variation in reflected light and therefore also in geometric albedo on the surface of a more or less spherical body. Such large variations are seen e.g. on the surface of Pluto (Bond albedo of A\,=\,0.2--0.9) and also on Charon (A\,=\,0.1--0.5), as revealed by New Horizons \citep{Buratti2017}. 

In the case of the coincident orbital/rotational axes configurations 
 (Cases~3a...3d) and 4, $\beta_{ss}$\,$\approx$\,51\degr) the dependence of the final solution on the thermal inertia chosen is relatively weak, and it cannot be well constrained by the far-infrared flux densities. 
The error bars of the observed flux densities allow acceptable solutions for these pole orientations as well: for  these cases 
we obtained  \deff\,=\,1224$\pm$55\,km and 1238$\pm$50\,km for the prograde solution for P\,=\,44.8\,h and 22.4\,h (3a and 3b), and \deff\,=\,1227$\pm$56\,km and 1241$\pm$50\,km for the retrograde cases with P\,=\,44.8\,h and 22.4\,h (3c and 3c), assuming Charon-like inertias. The similar sizes obtained show that (i) the effect of prograde/retrograde rotation is negligible for the thermal emission calculations and (ii) that the application of the slower/faster rotation introduces an uncertainty of $\sim$1\% in the estimated size and a corresponding $\sim$3\% uncertainty in volume and density (see below). Higher thermal inertias of a 'Pluto-like' surface provide \deff\, and \geomalb\, even closer to that of the Case~1 solution. {In these cases the observed $\Delta$m\,=\,0\fm09 light curve amplitude can be explained e.g. by a large, $\sim$60\degr-radius darker/brighter equatorial area with an albedo contrast of $\sim$17\% over the global value ($\Delta$\geomalb\,$\approx$\,2\% on the absolute scale), seen under the observed orbital inclination. }

Considering a tidally distorted ellipsoid compatible with the observed light curve (Cases~4 and 5a...5d) leads to somewhat different effective radii. In the equator-on case this leads to \deff\,=\,1549\,km, slightly ($\sim$1\%) larger than the corresponding spherical solution (Case~1). In the ellipsoidal pro-/retrograde equatorial satellite cases (5a...5b), however, the effective diameters obtained are typically $\sim$6\% smaller ($\sim$1160\,km) than in the spherical cases (3a...3d). The difference between the orbit solution and rotational period variations are, again, small, $\lesssim$1\%. 

In all the cases above $\chi^2_r$\,$\lessapprox$\,1, i.e. all these thermal emission solutions are acceptable for \orten.

\rm


\section{The density of \orten \label{sect:density}} 

To calculate the density from the mass of 1.75$\pm$0.07$\cdot$10$^{21}$\,kg 
we first used \deff\,=\,1535$^{+75}_{-225}$\,km, derived from radiometric models by \citet{Pal2016}, corresponding to our best-fit, Case-1 (equator-on) TPM solution. This provides an average density estimate of 0.92$_{-0.14}^{+0.46}$\,g\,cm$^{-3}$, assuming a spherical body 
Using the effective diameters from the {Case-3a-d} TPM solutions (satellite orbit in equatorial plane) the density is $\rho$\,=\,1.74$\pm$0.16\,g\,cm$^{-3}$.
\rm
The highest density, $\rho$\,=\,2.15$\pm$0.17\,g\,cm$^{-3}$, is obtained for the pole-on (Case-2) solution. { As was pointed out above, the triaxial ellipsoid cases (4, 5a-d) are a very unlikely option for a massive and slow rotating Kuiper belt object like \orten.}

\begin{figure*}[ht!]
\begin{center}
\includegraphics[width=0.5\textwidth,angle=270]{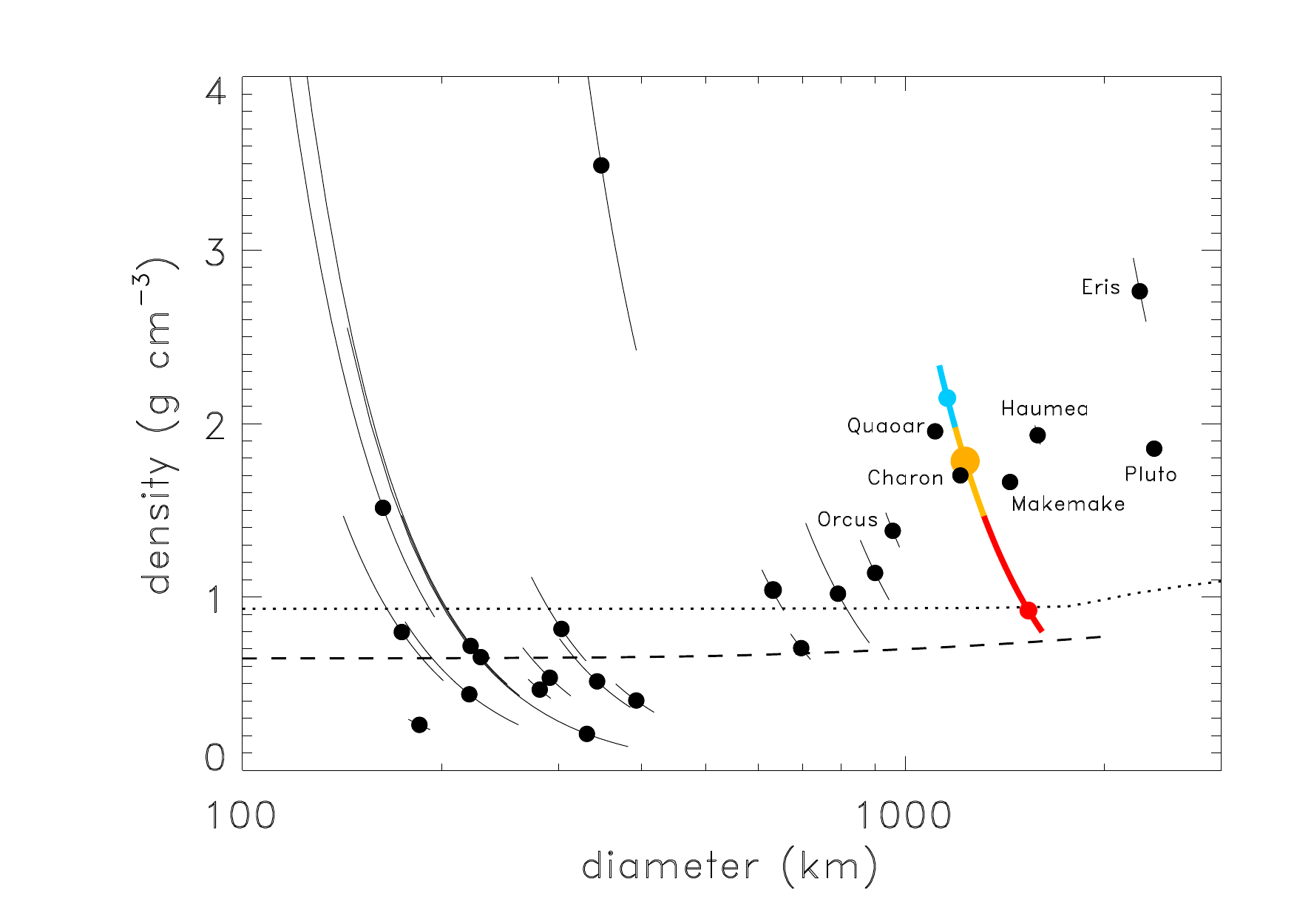}
\end{center}
\caption{Densities of trans-Neptunian objects as a function of their diameters. Majority of the data is taken from table~2 in \citet{Kovalenko}, but using the latest data for Haumea \citep{Ortiz2017}, Pluto and Charon \citet{Stern2015},  and for \ggg\, \citep{GGG}\rm. Colour symbols/arcs represent different densities obtained for \orten{} from the thermophysical model results: blue arc -- pole-on solution; orange arc -- satellite orbit in the equatorial plane of the primary; red arc -- equator-on solution. The dotted and dashed curves represent the density of a pure water ice sphere \citep{Lupo1979}, and the density expected from granular ice with self-compression \citep{McKinnon2005}, respectively. \label{fig:density}}
\end{figure*}

We compare the density of \orten{} with other trans-Neptunian object in Fig.~\ref{fig:density}. Considering the best-fit size (Case-1, red symbol and red arc in Fig.~\ref{fig:density}) the density of \orten{} is significantly lower than that of other objects with similar sizes, and rather similar to Kuiper belt object densities in the 500--1000\,km range. 
This would point to the highest ice / lowest rock fraction among the large Kuiper belt objects. The density of $\sim$0.92\,g\,cm$^{-3}$ is, however, consistent with the density of a pure water ice sphere (see Fig.~\ref{fig:density}). Such a low bulk density may also be a consequence of a core with a typical mixture of rock and ice (inside $\sim$50\% of the radius) and a highly porous mantle, where the low internal pressures may allow a porosity much higher (up to $\sim$50\%) than the residual porosities in the core ($\sim$10\%), as discussed e.g. for Quaoar in \citet{McKinnon2008}. 

 The Case-3a-d solutions ($\rho$\,=\,1.74$\pm$0.16\,g\,cm$^{-3}$), \rm put \orten\, in the range of densities defined by Charon, Haumea, Makemake, Orcus and Quaoar (orange symbol and arc in Fig.~\ref{fig:density}). 
Densities in this range are expected from the largest Kuiper belt objects if their moons are formed in collisions in which the primary retained its original composition and its primordial density \citep{Barr2016}. 

The high density obtained for the pole-on configuration (Case-2, $\rho$\,=\,2.15$\pm$0.17\,g\,cm$^{-3}$, blue symbol and arc in Fig.~\ref{fig:density}) is already in the range in which present day densities may have been caused by more energetic collisions, leading to a significant loss of ice. Again, this configuration is not very likely, due to the existence of a visible range light curve \citep{Pal2016}. 
The present accuracy of the radiometric size determination of \orten\, alone does not allow us to unambiguously choose between the possibilities presented above.  {However, considering all constraints, including the densities derived above, the most plausible solution for \orten\, seems to be a spherical shape with a single-peaked visible range light curve (P\,=\,22.4\,h) caused by albedo variegations, and co-planar primary equator and satellite orbit. This corresponds to the thermal emission solutions 3b and 3d.}

\section{Formation and tidal evolution}


To investigate the possible formation scenarios and the dependence of the tidal evolution on the basic properties of the system we considered a large number of configurations covering the possible size, density and structural properties of both the primary and the satellite, and estimated the tidal time scales and other parameters in a Monte-Carlo manner. 

Variables with known values are assumed to have a normal distribution with expectation value and standard deviation equal to their obtained values and uncertainties. These include the parameters of the satellite's orbit (semi-major axis, eccentricity) and also the properties that are directly derived from these parameters (system mass). The absolute magnitude of the primary and satellite are modelled in the same way (\absmag\,=\,2.34$\pm$0.01 and 6.93$\pm$0.15, respectively). 

In the case of variables with no known constraints we apply a feasible range of parameters and pick a specific value randomly. The geometric albedo of the satellite is chosen from \geomalb\,=\,0.01--1.0, and the effective radius is obtained from the absolute magnitude and \geomalb\, assumed. 
We used the effective diameter range of 1126 to 1610\,km for the main body, as given in Sect.~\ref{Sect:Radiometry}. 

\begin{figure*}[ht!]
\begin{center}
\hbox{
\includegraphics[width=0.48\textwidth]{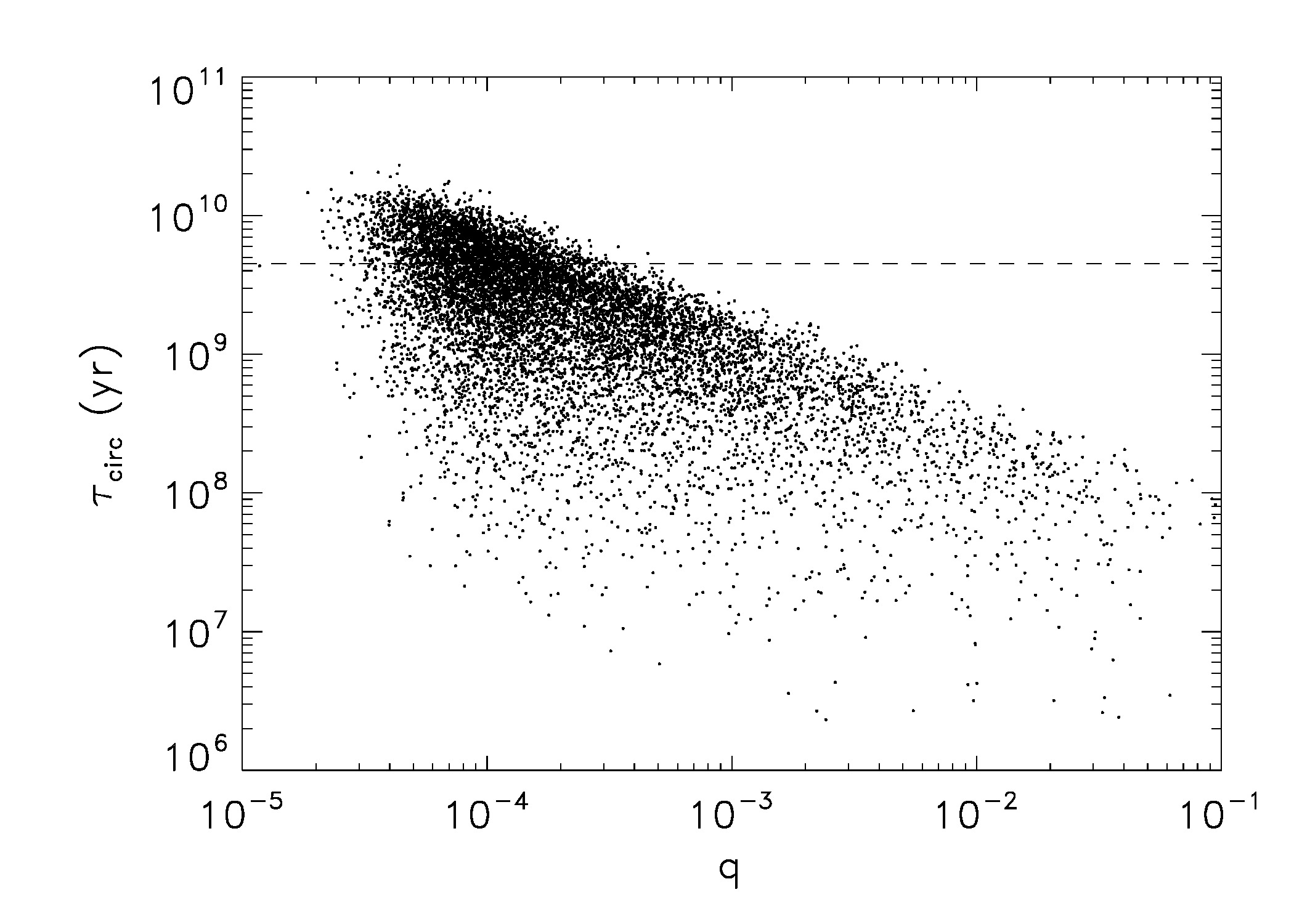}
\includegraphics[width=0.48\textwidth]{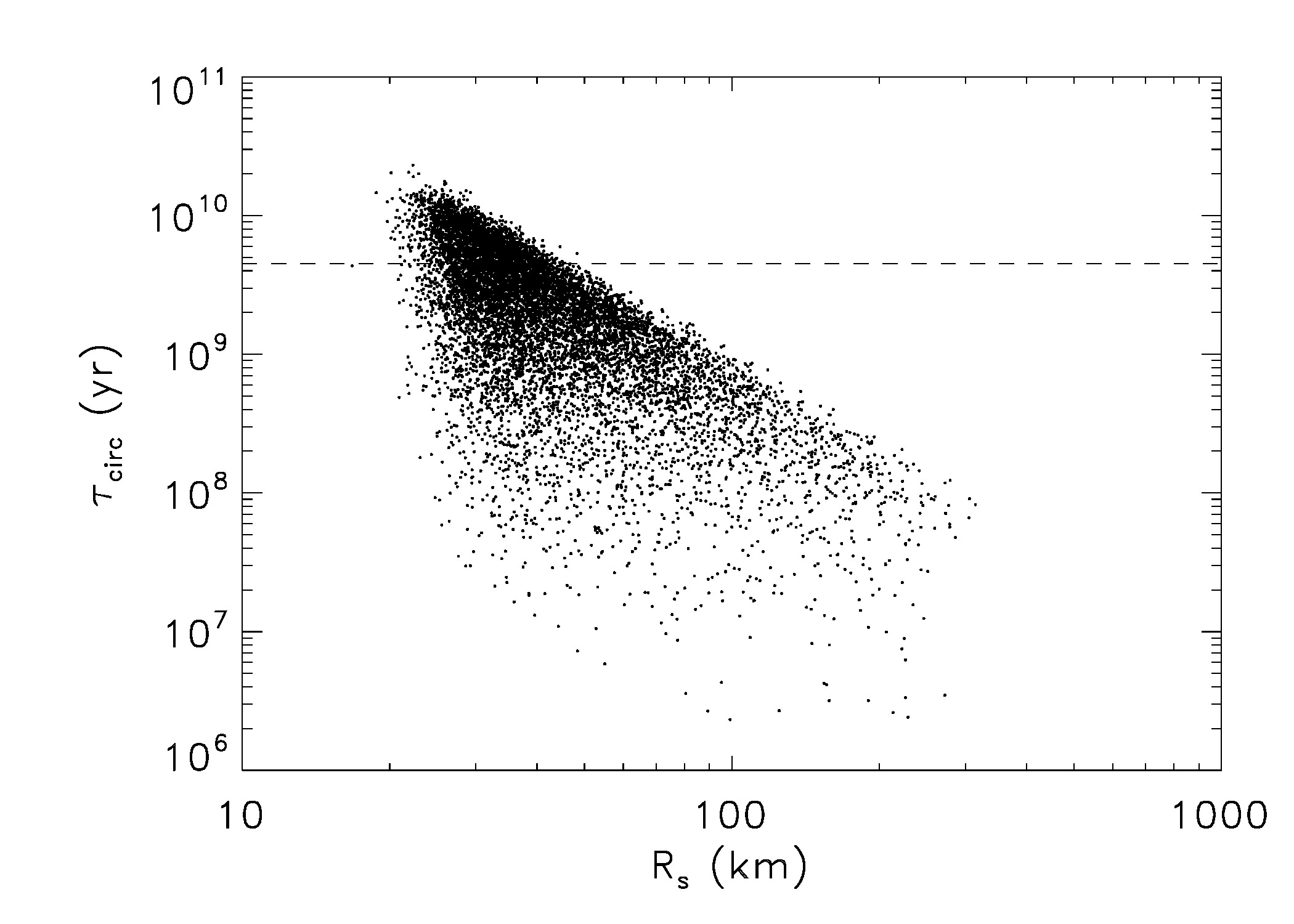}}
\hbox{
\includegraphics[width=0.48\textwidth]{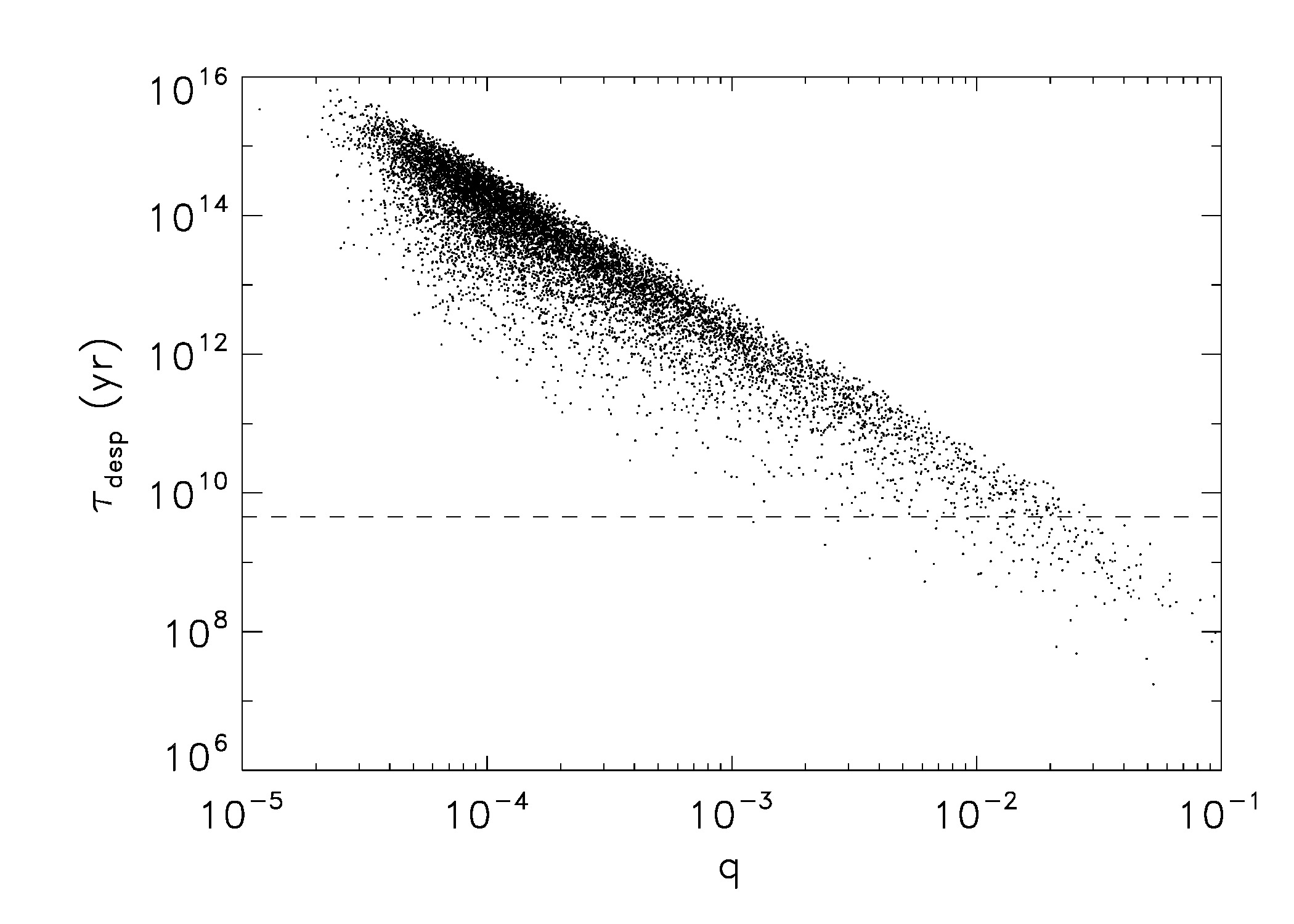}
\includegraphics[width=0.48\textwidth]{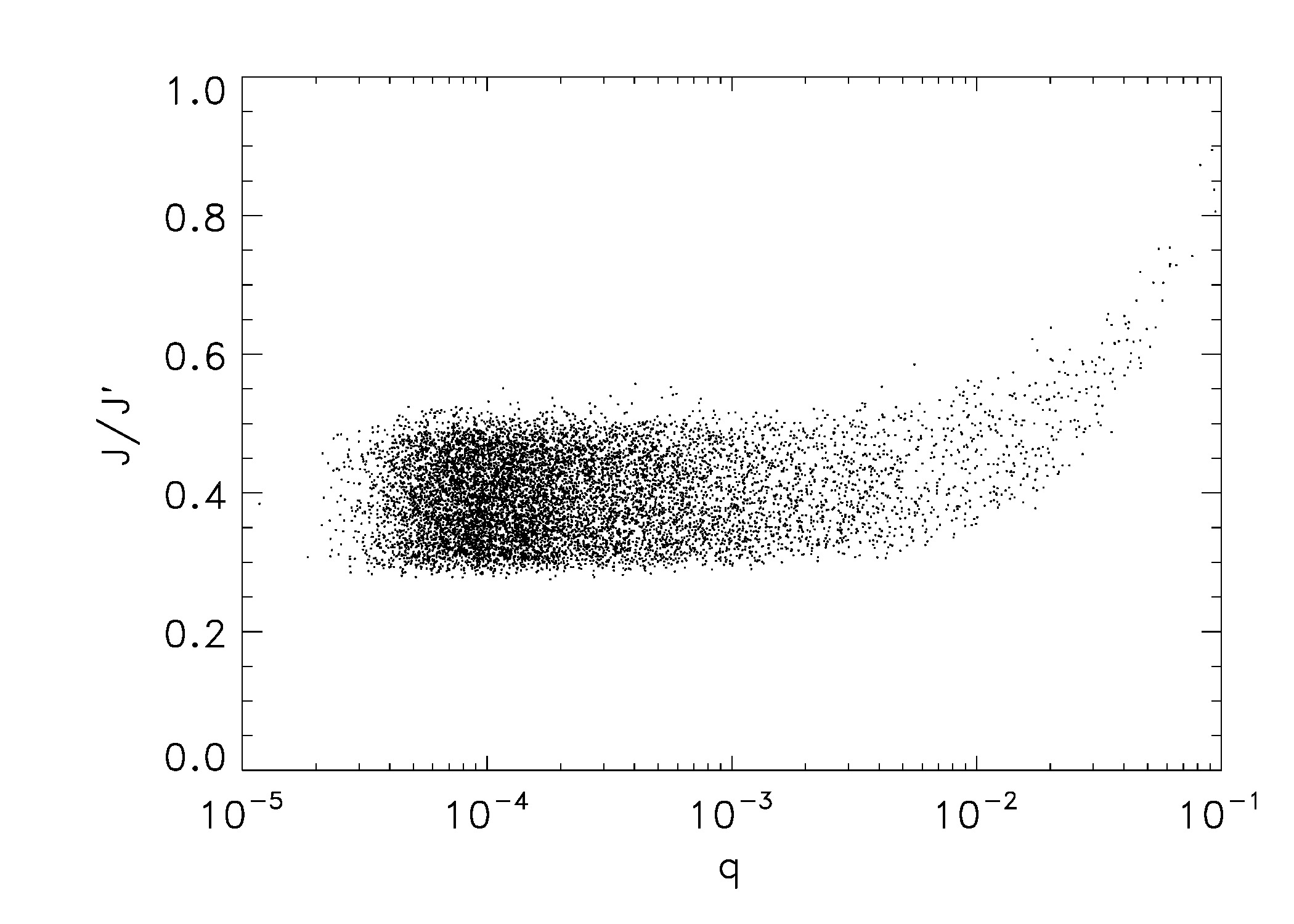}}
\end{center}
\caption{{\it Top-left}: orbit circularization timescales, $\tau_{\mathrm circ}$  vs. satellite to primary mass ratio (q);
{\it Top-right}: $\tau_{\mathrm circ}$ vs. the effective radius of the satellite (R$_S$); 
{\it Bottom-left}: Despinning timescale vs. q of the primary; 
{\it Bottom-right}: Normalized angular momentum (J/J') vs. q. 
The dashed horizontal line represents the age of the Solar system on those figures where timescales are plotted. 
\label{fig:tau_circ}}
\end{figure*}

We estimated the orbit circularization timescale ($\tau_{circ}$) of the system following \citet[][eq.~8]{Noll2008}, and a tidal dissipation factor Q was assumed in the range of 10-500 \citep[see e.g.][]{Goldreich1966,Farinella1979}. The obtained $\tau_{circ}$ values are plotted in Fig.~\ref{fig:tau_circ}. 
As the system has a notable eccentricity $\tau_{circ}$ should at least be larger than the age of the Solar system, 
assuming that the binary system formed 4.5\,Gyr ago, either by major impact or capture.
The only configurations that fulfill this requirement are those for which q\,$<$4$\cdot$10$^{-4}$ or R$_S$\,=\,18--50\,km, corresponding to geometric albedos of \geomalb\,=\,1.0--0.2. For these configurations the tidal factor must also be large, Q\,$>$\,100 in all cases, in agreement with that found e.g. for the icy satellites of the giant planets \citep{Goldreich1966}. Small tidal factors would result in a faster orbital evolution, not compatible with the observed moderate eccentricity. 

Similarly, the spin-locking or despinning timescale, $\tau_{desp}$, can also be estimated for both the primary and the satellite, using eq.~9 in \citet{Noll2008}. The $\tau_{desp}$ is below 10$^7$\,yr for all of our model configurations, i.e. the rotation of the satellite is almost certainly tidally locked. For the primary, however, these timescales are much longer. $\tau_{desp}$ is below 4.5\,Gyr only when q\,$>$\,2$\cdot$10$^{-2}$. This large q would, on the other hand, lead to a fast circularization of the orbit that obviously did not happen. $\tau_{desp}$\,$>$\,10$^{12}$\,yr for the system parameters allowed by the observed eccentricity.

The tidal despinning timescales derived above strongly suggest that the observed lightcurve can be attributed to the primary.
As suggested in \citet{Pal2016}, the observed rotation period may be the orbital period of two nearby, tidally locked bodies. Knowing the system mass we can calculate the separation of such a semi-contact binary.
In this case the separation of the two bodies would be 4250\,km or 67\,mas. 
We investigated the co-added images of each observational epoch to identify any deviation from a single-source point spread function (PSF). Model PSFs were created using the TinyTim \citep{Krist2010} software 
applied on two point sources with the expected separation and a range of relative brightnesses (1:2 to 1:20). A comparison of the model and observed PSFs show no signs of notable distortion at any of the 2017 epochs, down to the brightness ratio of 1:10 at which a double system would still be detectable. 




We calculated the normalized angular momentum of the system \citep[J/J'][]{Noll2008}, considering the combined spin and orbital angular momentum, for a wide range of system parameters. 
The dependence of J/J' on the primary to satellite mass ratio 
q is presented in Fig.~\ref{fig:tau_circ}. For smaller q values J/J' converges to $\sim$0.4 (very close to that of Pluto-Charon), and despite that it decreases towards larger satellite masses it remains J/J'\,$<$\,0.8 even for the largest q-s. As discussed in \citet{Noll2008} binary systems produced by single collisions should have J/J'\,$<$\,0.8, a condition that is fulfilled by the \orten{} system.


The evolution of the satellite orbit may be governed by the Kozai mechanism in the case of \orten, due to perturbations by the Sun \citep[for a detailed discussion of the Kozai mechanism and its implications for trans-Neptunian binaries, see][]{PN09}. 
The inclination of the satellite orbit to the heliocentric orbit, $i_h$, is  51\fdg83 \rm (prograde) or  129\fdg05 \rm (retrograde). Because, in the quadrupole approximation, $\sqrt{1-e^2}$\,$\cos i_h$ is conserved \citep{Naoz2016}, the possible ranges of eccentricity and inclination that the system may take are  0\,$\leq$\,$e$\,$\leq$\,0.65 and 39$\degr$\,$\leq$\,$i_h$\,$\leq$\,54$\degr$ for prograde, and 0\,$\leq$\,$e$\,$\leq$\,0.63 and 127$\degr$\,$\leq$\,$i_h$\,$\leq$\,141$\degr$ for retrograde orbit, with an associated timescale of $\sim$2$\cdot 10^6$\,yr \citep[see eq.~1 in][]{PN09}.  Assuming that the present orbit is a consequence of the Kozai mechanism, and the system originally had an eccentricity close to zero, the initial inclination should have been  $i_0$\,$\approx$\,54\degr. \rm  As the orbit of the satellite is not circularized we may also put constraints on the strength of the combined Kozai and tidal effects \citep{PN09}. For an initial inclination of  $i_0$\,$\approx$\,54\degr{} \rm a system with orbital semi-major axis to characteristic tidal distance ratio of a/r$_c$\,$\leq$\,1.5 should have evolved to e\,$\approx$\,0 by now, i.e. a nearly circular orbit. The a/r$_c$ value depends primarily on q, and e\,$\gneqq$\,0 requires q\,$\leq$\,5$\cdot$10$^{-3}$, obtained using the same approach as discussed in the case of the tidal timescales. This upper limit for q is in agreement with those obtained from other tidal timescale calculations above.   


Irregularly shaped bodies have higher order terms in their gravitational potential which may dominate over the solar tides, the latter one responsible for the Kozai oscillations \citep{Nicholson2008,Grundy2011}. The most important quadrupole term is related to the flattening, $\epsilon$, of the main body through the J$_2$ dynamic form factor. 
Assuming a Maclaurin ellipsoid -- flattening due to rotation of a body with homogeneous internal density distribution -- we can estimate the flattening of \orten, following \citet{Plummer1919}. This results in flattening values of 0.0026\,$\le$\,$\epsilon$\,$\le$\,0.0118, assuming a range of sizes and densities as in the calculation of the other dynamical timescale above. The corresponding form factors are in the range of 0.0001\,$\le$\,J$_2$\,$\le$\,0.04 \citep[e.g.][]{Essen}. 
We calculated the critical semi-major axis $a_c$ between the oblateness-dominated and solar-tide-dominated dynamics \citep[e.g. eq.~3 in][]{Nicholson2008} and obtained 0.35\,$\le$\,$a$/$a_c$\,$\le$\,0.49, where $a$ is the semi-major axis of the satellite orbit in the \orten{} system. This suggests that that dynamics of \orten's satellite should be governed by the oblateness of the primary, and not by solar tides, at least based on the present orbit. The associated  precession timescales are 5.4$\cdot$10$^4$\,$\le$\,$\tau_p$\,$\le$\,4.2$\cdot$10$^5$\,yr for the prograde and 1.8$\cdot$10$^4$\,$\le$\,$\tau_p$\,$\le$\,1.4$\cdot$10$^5$\,yr for the retrograde case.
\citet{Grundy2011} obtained $a$/$a_c$ ratios for 17 trans-Neptunian binary systems, and in this sample there are only three systems (1999\,OJ$_4$, (123509)\,2000\,WK$_{183}$, (66652)\,Borasisi) where the calculated $a$/$a_c$ ratio is so low that that system is almost certainly in the oblateness-dominated regime. For \orten\, an oblateness-dominated dynamics should have lead to a circularized orbit. 
 
\section{Conclusions \label{Conclusions}}

In most of the calculations above \orten's satellite must be small in order to keep the satellite orbit from circularization during the lifetime of the solar system. While other mechanisms may play a role and increase the eccentricity from a small value to the presently observed one, a small satellite (R$_s$\,$<$\,50\,km) with a relatively bright surface (\geomalb\,$>$\,0.2) would be consistent with all possible evolutionary scenarios. Among the largest Kuiper belt objects Quaoar and Haumea have similarly small satellites and low relative mass ratios \citep{Barr2016}; the small satellites of Pluto also show high albedo values \citep{Weaver2016}. 
With respect to orbital characteristics, \orten's satellite is similar to Weywot that also has an eccentric orbit around Quaoar \citep[e\,$\approx$\,0.14,][]{Fraser2013}. An even smaller satellite, with a mass ratio of q\,$\le$\,5$\cdot$10$^{-3}$, however, is not likely to have been able to slow down the rotation of the primary to the present $\sim$45\,h, if it originally had a rotation period typical for a Kuiper belt object \citep[8.6\,h][]{Thirouin2014}. 


The present accuracy of the radiometric size determination of \orten\, does not allow us to unambiguously choose between the possible densities. The solution depends mainly on the orientation of the spin axis -- a larger subsolar latitude, $\beta_{ss}$, leads to a smaller size and a higher density. Due to current large heliocentric distance and the cold surface temperatures measurements in the mid-infrared range ($\sim$10--25$\mu$m) would not significantly improve the radiometric models \citep[see the thermal emission modelling in][]{Pal2016}. An eccentricity $\epsilon$\,$\approx$\,0 would make the equatorial plane 
satellite orbit significantly more likely, but the present orbit does not allow us to draw a definite conclusion on the relative positions of the two planes. 
Future occultation measurements and/or direct imaging e.g. by the James Webb Space Telescope may be able to reveal the true size and decide on the density.

Our simple dynamical considerations could not reveal the mechanism that could have lead to the present orbit. In one possible scenario the satellite of \orten\, could initially be a captured satellite in a distant orbit, where the Kozai mechanism pumped the 
eccentricity until tidal evolution took over, and finally this tidal dissipation
shrank the orbit to an oblateness-dominated regime. In this regime the non-circularized orbit may be explained, if the mass of the satellite is really small, as it is indicated by the dynamical timescale calculations above. More complex scenarios, like the involvement of spin-orbit resonances may also lead to the present orbit. A more detailed analysis of the dynamics of \orten's satellite and its possible origin and evolution will be performed in a forthcoming paper. 


\section*{Acknowledgements}

Data presented in this paper were obtained from the Mikulski Archive for Space Telescopes (MAST). STScI is operated by the Association of Universities for Research in Astronomy, Inc., under NASA contract NAS5-26555. This work is based in part on NASA/ESA Hubble
Space Telescope program 15207. Support for this program was provided by
NASA through grants from the Space Telescope Science Institute
(STScI). Support for MAST for non-HST data is provided by the NASA Office of Space Science via grant NNX09AF08G and by other grants and contracts. The research leading to these results has received funding from the European Union’s Horizon 2020 Research and Innovation Programme, under Grant Agreement no 687378; from the K-125015 and GINOP-2.3.2-15-2016-00003 grants of the National Research, Development and Innovation Office (NKFIH, Hungary). MES was supported by Gemini Observatory which is operated by the Association of Universities for Research in Astronomy, Inc., on behalf of the international Gemini partnership of Argentina, Brazil, Canada, Chile, and the United States of America. 


\end{document}